\documentclass[final,1p,times]{elsarticle}
\usepackage{graphics}
\usepackage{graphicx}
\usepackage{epstopdf}
\usepackage{ifpdf}
\usepackage{amssymb}
\usepackage{amsthm}
\usepackage{lineno}
\usepackage{dcolumn}
\usepackage{bm}
\usepackage{epsfig}
\usepackage{ifpdf}
\usepackage{amsthm}
\usepackage{subfig}

\newdefinition{rmk}{Remark}
\newproof{pf}{Proof}
\newproof{pot}{Proof of Theorem \ref{thm2}}
\journal{none}

\begin{document}
\begin{frontmatter}
\title{Fermionic and scalar fields as sources of interacting dark matter-dark energy}
\author[ucv]{Samuel Lepe}
\ead{slepe@ucv.cl}
\author[dcf]{Javier Lorca}
\ead{j.lorca@ufro.cl}
\author[dcf]{Francisco Pe\~na}
\ead{fcampos@ufro.cl}
\author[dcf]{Yerko V\'asquez}
\ead{yvasquez@ufro.cl}
\address[ucv]{Instituto de F\'\i sica, Facultad de Ciencias, Pontificia Universidad Cat\'olica de Valpara\'\i so, Casilla 4059, Valpara\'\i so, Chile}
\address[dcf]{Departamento de Ciencias F\'\i sicas, Facultad de Ingenier\'\i a, Ciencias y
Administraci\'on, Universidad de La Frontera, Avda. Francisco Salazar 01145,
Casilla 54-D Temuco, Chile.}

\begin{abstract}
From a variational action with non-minimal coupling with a scalar field and classical scalar and fermionic interaction, cosmological field equations can be obtained. Imposing a FLRW metric the equations lead directly to a cosmological model consisting of two interacting fluids, where the scalar field fluid is interpreted as dark energy and the fermionic field fluid is interpreted as dark matter. Several cases were studied analytically and numerically. An important feature of the non-minimal coupling is that it allows crossing the barrier from a quintessence to phantom behavior. The insensitivity of the solutions to one of the parameters of the model permits it to find an almost analytical solution for the cosmological constant type of universe.
\end{abstract}

\begin{keyword}
Cosmology \sep Non-minimal coupling \sep Fermions \sep Scalars \sep Dark Energy
\end{keyword}
\end{frontmatter}

\linenumbers

\section{ Introduction}

In this paper we have considered fermionic and scalar fields as sources of dark matter and dark energy, respectively. Two types of interaction have been modeled, an interaction between the scalar field and fermionic field of the Yukawa type and a non-minimal interaction between the scalar and gravitational field. Recent observations suggest a equation of state of the type $\omega<-1$ \cite{komatsu}, which makes possible to consider cosmological models admitting dynamical equations of state allowing to cross the barrier $\omega=-1$. A minimal coupling to the scalar field is not enough to achieve a barrier cross to the phantom zone, however a non-minimal coupling allows this feature. This type of coupling has been studied in inflationary scenarios and in Grand Unified Theories (GUT) \cite{Buchbinder, Elizalde, Muta}.

We have studied the implications that these interactions have, specially non-minimal coupling, regarding the cosmic evolution of the model. Two Ansatz have been used and explored for the energy density transmission which are proportional to each energy density. We have found solutions to the field equations in a FRW flat universe. For these solutions the bilinear scalar $S=\bar{\psi}\psi$ shows a monotonic decrease in an expanding universe, this is, proportional with $a^{-3}$. This fact does not imply spinor energy density to have the same behavior. By redefining the energy densities, a new equivalent and symmetric system of equations is found which manifestly has a redefined conserved energy momentum tensor. Inevitably, these redefinitions introduce two critical values to the scalar field, which are impossible barriers for the field to cross over, therefore, the cosmological solutions are valid between these two values \cite{faraoni}.

For the non-minimal parameter $\xi \neq 0$ and the energy density transfer function $\hat{Q}$ proportional to $\hat{\rho_{\psi}}$, we found a de Sitter university evolution coming from a decelerated phase to an accelerated phase. There is no crossing of the barrier $\omega=-1$, except for a large interaction parameter $\lambda$ which doesn't have a clear physical ground. For the redefined energy density transfer function $\hat{Q}$ proportional to $\hat{\rho_{\phi}}$ phantom solutions are obtained, and when the scale factor is near $a_{c}$  the Hubble parameter and energy densities diverge, finding a future singularity.
To consider a non-minimal coupling, allows crossing the barrier $\omega=-1$, where there is a transient from a dark matter dominated universe to a dark energy dominated universe, quintessence, and then a phantom evolution.

We have a future singularity, in which the Hubble parameter, redefined energy densities and state equations diverge for a finite size, this future singularity correspond to a type III \cite{nojiri}.

In the first section we present the Lagrangian model and field equations derived from it, and are particularized for the metric FRW. In the second section the cosmological field equations are found from the field equations derived in the first section. Two cases are studied: minimal and non-minimal coupling. In the third section the cosmological solutions are analyzed for two types of Ansatz interacting function. In the last section a discussion of the results are presented along with the curves derived from the cosmological field equations. A quasi analytical solution for these systems is presented for which a more detailed derivation is attached at the Appendix section.

\section{Formalism and field equations}

In this section, a brief description of the techniques used to include fermionic and scalars sources in mutual interaction with the Einstein theory of gravitation are presented \cite{delamacorra,micheletti,ribas}. Due to the fact that the gauge group of General Relativity does not admit a spinor representation the tetrad formalism is invoked. Following the general covariance principle, a connection between the tetrad and the metric tensor $g_{\mu \nu }$ is established through the relation
\begin{eqnarray}  \label{metrictetrad}
g_{\mu\nu}=e_{\mu}^{a}e_{\nu}^{b}\eta_{ab},\;\;\; a,b=0,1,2,3
\end{eqnarray}
where $e_{\mu}^{a}$ denotes the tetrad or "vierbein" and $\eta_{ab}$ is the Minkowski metric tensor. Here and after, Latin indices refer to the local inertial frame whereas Greek indices to the bundle space $\mathcal{M}$. The main objective of this work is to describe the behavior of fermions $\psi:\mathcal{M} \rightarrow \mathbb{R}^4$ and scalars $\phi: \mathcal{M} \rightarrow \mathbb{R}$ with self-interacting potential density $V(\bar{\psi},\psi):\mathcal{M} \rightarrow \mathbb{R}$ and $U(\phi):\mathcal{M}\rightarrow \mathbb{R}$ in presence of a gravitational field. The dynamics between fermions and scalars fields will be represented on the Lagrangian by a Yukawa type interaction through the map $f\left( \phi \right):\mathcal{M}\rightarrow \mathbb{R}$. The action for this system is,
\begin{eqnarray}  \label{action}
\mathcal{S}(g,\psi,\bar{\psi},\phi)=\int d^{4}x \sqrt{-g}\mathcal{L},
\end{eqnarray}
where
\begin{eqnarray}  \label{lagrangiana}
\mathcal{L}=\frac{{1}}{{2}}(1-\xi\phi^2)R+\frac{i}{2}\left(\bar{\psi}{\Gamma^{\mu}}{\bigtriangledown_{\mu}}\psi- {\bigtriangledown_{\mu}}{\bar{\psi}}\Gamma^{\mu}\psi\right)-m{\bar{\psi}}\psi +  \nonumber \\
- V(\bar{\psi},\psi)+\frac{1}{2}\partial_{\mu}\phi\partial^{\mu}\phi - U(\phi)-{\bar{\psi}}f(\phi)\psi,
\end{eqnarray}
remarking that spinors are treated here as classical commuting fields and $\xi \in \mathbb{R}$ is a parameter for a non-minimal coupling between gravitation and scalar fields \cite{nozari,faraoni}. Lagrangian density (\ref{lagrangiana}) uses natural units, i. e. $8 \pi G = c = \hbar =1$, also $m$ is the bare fermionic mass, $\bar{\psi} =\psi^{\dag}\gamma^{0}$ and $R$ denotes the scalar of curvature. The Dirac matrices can be generalized to curved space through the definition $\Gamma^{\mu}=e^{\mu}_{a} \gamma^{a}$ satisfying $\{\Gamma^{\mu},\Gamma^{\nu}\}=2g^{\mu\nu}$. The covariant derivative is then ${\bigtriangledown_{\mu}}\psi=\partial_{\mu} \psi-\Omega_{\mu}\psi$  and   ${\bigtriangledown_{\mu}}\bar{\psi}=\partial_{\mu} \bar{\psi}+\bar{\psi}\Omega_{\mu}$, where the spin connection $\Omega_{\mu}$
is given by
\begin{equation}\label{spinconnection}
\Omega_{\mu}=-\frac{1}{4}g_{\mu\nu}\left[\Gamma^{\nu}_{\sigma\lambda}-
e^{\nu}_{b}(\partial_{\sigma}e^{b}_{\lambda})\right]\gamma^{\sigma}\gamma^{%
\lambda},
\end{equation}
with $\Gamma^{\nu}_{\sigma\lambda}$ denoting the Christoffel symbols.

The field equations are obtained by varying the total action (\ref{action}) with respect to the tetrad, spinor field and scalar field, respectively. By defining $\alpha =1-\xi \phi ^{2}$,  the following equations are obtained:
\begin{eqnarray}
\label{einsteinequations}R_{\mu \nu }-\frac{1}{2}g_{\mu \nu }R &=&\alpha ^{-1}T_{\mu \nu },\\
\label{phiequations}D_{\mu }D^{\mu }\phi+\frac{\partial U(\phi )}{\partial \phi }+\bar{\psi}\frac{\partial f(\phi )}{\partial \phi }\psi &=& \frac{\partial \alpha \left( \phi \right)}{\partial \phi} R, \\
\label{psiequations}i\Gamma ^{\mu }\nabla _{\mu }\psi -(m+f(\phi ))\psi &=&\frac{\partial V(\bar{%
\psi}\psi )}{\partial {\bar{\psi}}},\\
\label{barpsiequations} i\nabla _{\mu }\bar{\psi}\Gamma ^{\mu }+(m+f(\phi ))\bar{\psi} &=&-\frac{%
\partial V(\bar{\psi}\psi )}{\partial {\psi }},
\end{eqnarray}
where $T_{\mu \nu }=T_{\mu \nu }^{\phi }+T_{\mu \nu }^{D}+T_{\mu \nu }^{int}$ and
\begin{eqnarray}
\label{tinteraction} T_{\mu \nu }^{\phi } &=&-\partial _{\mu }\phi \partial _{\nu }\phi +\frac{1}{2}g_{\mu \nu }\partial _{\rho }\phi \partial ^{\rho }\phi -g_{\mu \nu}U(\phi )-g_{\mu \nu }\square \alpha (\phi )+\partial _{\mu }\partial _{\nu}\alpha (\phi ),  \\
\nonumber T_{\mu \nu }^{D} &=&\frac{i}{4}\left[ \bar{\psi}\Gamma ^{\mu }\nabla ^{\nu
}\psi +\bar{\psi}\Gamma ^{\nu }\nabla ^{\mu }\psi -\nabla ^{\nu }\bar{\psi}\Gamma ^{\mu }\psi -\nabla ^{\mu }\bar{\psi}\Gamma ^{\nu }\psi \right]
-g_{\mu \nu }\mathcal{L}_{D}, \\
\nonumber T_{\mu \nu }^{int} &=&-g_{\mu \nu }\bar{\psi}f(\phi )\psi ,
\end{eqnarray}
and $\mathcal{L}_{D}=\frac{i}{2}\left( \bar{\psi}{\Gamma ^{\mu }}{\bigtriangledown _{\mu }}\psi -{\bigtriangledown _{\mu }}{\bar{\psi}}\Gamma^{\mu }\psi \right) -m{\bar{\psi}}\psi -V(\bar{\psi},\psi )$ is the Dirac Lagrangian.

In the following, we consider a FLRW flat universe described by the metric
\begin{equation}  \label{FRWmetric}
ds^{2}=dt^{2}-a(t)^{2}[dx^{2}+dy^{2}+dz^{2}],
\end{equation}
where $a(t)$ denotes the cosmic scale factor. According to the metric (\ref{FRWmetric}), the tetrad components read
\begin{equation}  \label{tetrad}
e^{\mu}_{0}=\delta^{\mu}_{0}\;\;,\;\; e^{\mu}_{i}=\frac{1}{a(t)}%
\delta^{\mu}_{i},
\end{equation}
and Dirac matrices become
\begin{equation}  \label{Diracmatrices}
\Gamma^{0}=\gamma^{0}\;\;,\;\;\Gamma^{i}=\frac{1}{a(t)}\gamma^{i},
\end{equation}
from which the spin connection components are obtained, yielding
\begin{equation}  \label{spinconnection}
\Omega_{0}=0\;\;,\;\;\Omega_{i}=\frac{1}{2}\dot{a}(t)\gamma_{i}\gamma_{0},
\end{equation}
where a dot for the time derivative have been introduced.

We consider now that the fields are homogenous and isotropic. This is based on the observational fact that on a cosmological scale higher than $300 Mpc$ the fields appear to be independent of the spatial coordinates in a post inflation evolution \cite{komatsu}. In order to study this model exhaustively we will consider the cases of a minimal coupling $\xi=0$ and a non-minimal coupling $\xi \neq 0$.

Fermion field equations (\ref{psiequations}) and (\ref{barpsiequations}) become
\begin{eqnarray}
\label{psipointequation} \dot{\psi}+\frac{3}{2}H\psi &=&-i(m+f\left(\phi\right))\gamma ^{0}\psi -i\gamma ^{0}\frac{%
\partial V}{\partial \bar{\psi}},\\
\dot{\bar{\psi}}+\frac{3}{2}H\bar{\psi} &=&i(m+f\left(\phi\right))\bar{\psi}\gamma ^{0}+i%
\label{barpsipointequation}\frac{\partial V}{\partial \psi }\gamma ^{0}\;\;,
\end{eqnarray}%
where $H=H(t)=\frac{\dot{a}(t)}{a(t)}$ is the Hubble parameter. From equations (\ref{psipointequation}) and (\ref{barpsipointequation}) the following relation is obtained:
\begin{equation}\label{fermionescalar}
\frac{d}{dt}\left( \bar{\psi}\psi \right) + 3H\left( \bar{\psi}\psi \right)= i \left( \frac{\partial V}{\partial \psi} \gamma^0 \psi - \bar \psi \gamma^0 \frac{\partial V}{\partial \bar \psi} \right).
\end{equation}
Note that considering a self interacting potential of the form $V\left(\bar \psi, \psi \right)=V \left( \bar \psi \Gamma \psi \right)$ turn null the right side of equation (\ref{fermionescalar}); in fact, with this assumption equation (\ref{fermionescalar}) can be immediately integrated, yielding
\begin{equation}\label{fermionscalarminimal}
S=S_{0}\left( \frac{a_{0}}{a}\right) ^{3},
\end{equation}
where $S=\bar{\psi}\psi$ has been defined.

\section{Cosmological Field Equations}

We will consider two cases, a minimal coupling $\xi=0$ $(\alpha = 1)$ and a non-minimal coupling $\xi \neq 0$.

\subsection{Minimal coupling $\xi =0$}

In a minimal coupling $\alpha = 1$ which simplifies considerably the field equations (\ref{einsteinequations} - \ref{barpsiequations}). It is known that the Einstein field equations (\ref{einsteinequations}) fulfill
\begin{eqnarray}
\label{accelerationequationminimal}\frac{\ddot{a}}{a} &=&-\frac{1}{6}[\rho _{\phi }+\rho _{\psi }+3(p_{\phi
}+p_{\psi })],\\
\label{friedmanequationminimal}3H^{2} &=&\rho _{\phi }+\rho _{\psi },
\end{eqnarray}
known as the acceleration (\ref{accelerationequationminimal}) and Friedman's equations (\ref{friedmanequationminimal}). On the other hand, the scalar field equation (\ref{phiequations}) can be written
\begin{equation} \label{barphipointequation}
\ddot{\phi}+3H\dot{\phi}=-\frac{\partial U}{\partial \phi }-\bar{\psi}
\frac{\partial f}{\partial \phi }\psi \;\;.
\end{equation}
allowing the following identifications
\begin{eqnarray}
\label{rho_phi_minimal} \rho _{\phi } &=&\frac{1}{2}\dot{\phi}^{2}+U(\phi ),\\
\label{rho_psi_minimal} \rho _{\psi } &=&\bar{\psi}(m+f\left(\phi\right))\psi +V(\bar{\psi},\psi ),\\
\label{p_phi_minimal} p_{\phi } &=&\frac{1}{2}\dot{\phi}^{2}-U(\phi ),\\
\label{p_psi_minimal} p_{\psi } &=& \frac{1}{2}\bar{\psi}\frac{\partial V}{\partial \bar{\psi}}+
\frac{1}{2}\frac{\partial V}{\partial \psi }\psi -V(\bar{\psi},\psi ).
\end{eqnarray}

From (\ref{rho_phi_minimal} - \ref{p_psi_minimal}) regarding the fact that $\dot V = -3HS\frac{dV}{dS} $ it is straightforward to obtain
\begin{eqnarray}
\label{dot_rho_phi_minimal}\dot{\rho}_{\phi }+3H(\rho _{\phi }+p_{\phi })&=&-Q,\\
\label{dot_rho_psi_minimal}\dot{\rho}_{\psi }+3H(\rho _{\psi }+p_{\psi })&=&Q,
\end{eqnarray}
where $Q=\bar{\psi}\frac{\partial f}{\partial \phi }\psi \dot{\phi}$ is recognized directly from the field equations as the interaction function. The equation of state for each field is defined by $\omega_\chi = \frac{p_\chi}{\rho_\chi}$, where $\chi=\phi$ or $\chi=\psi$.

This last remark allows to interpret \textit{fermionic fields as sources of dark matter} and \textit{bosonic fields as sources of dark energy}. We will see however that this same identification can be made in the case of a non-minimal coupling.

\subsection{Non-minimal coupling $\xi \neq 0$}

In the case of non-minimal coupling, assuming $\omega _{\psi }=0$, i.e. a dust type solution, equations (\ref{einsteinequations} - \ref{barpsiequations}) take the following form
\begin{eqnarray}
\label{SISTEMA1}\dot{\rho}_{\phi }+3H\rho _{\phi }\left( 1+\omega _{\phi }\right)  &=&-\frac{2\xi \phi \dot{\phi}\left( \rho _{\phi }+\rho _{\psi }\right) }{1-\xi \phi^{2}}-Q,  \\
\label{SISTEMA2}\dot{\rho}_{\psi }+3H\rho _{\psi } &=& Q, \\
\label{SISTEMA3}Q &=&\beta S_{0}a^{-3}\dot{\phi},
\end{eqnarray}
equations (\ref{SISTEMA1}) and (\ref{SISTEMA2}) imply a non conservation of the energy momentum tensor $T_{\mu \nu}$, however it will be shown below that there exists a redefined energy-momentum tensor $\hat T_{\mu \nu}$ that is conserved. Equation (\ref{SISTEMA3}) is easily obtained by considering the Yukawa type of interaction $f \left( \phi \right) = \beta \phi$ and equation (\ref{fermionscalarminimal}) . The last two equations can be combined to yield an exact differential in terms of the scale factor $a$
\begin{eqnarray}
\nonumber \frac{d}{da}\left(\frac{\rho_\psi a^3}{\beta S_0} - \phi \right) &=& 0, \\
\label{phisolution} \phi =\frac{\rho _{\psi }}{\beta S_{0}}a^{3} + c.
\end{eqnarray}
where the integration constant $c$ will be shown to be zero.

As it can be seen, there exist a singularity point $a=a_{c}$ when $1-\xi \phi^2 \left(a_{c}\right)=0$ or equivalently when $\rho _{\psi }\left( a_{c}\right) =\frac{\beta S_{0}}{\sqrt{\xi }}\frac{1}{a_{c}^{3}}$ with $\xi >0$ in which the fields and the gradient of the fields diverge, therefore the model is valid only for $|\phi| < \phi \left( a_{c} \right)$ or $|\phi| > \phi \left( a_{c} \right)$. For $\xi >0$ a restriction for the field equations and a limited class of solutions is obtained \cite{faraoni}.

Symmetry of the last system of equations can be restored by the following procedure. In the non-minimal coupling case, equations (\ref{accelerationequationminimal}) and (\ref{friedmanequationminimal}) can be generalized to
\begin{eqnarray}
\label{accelerationequationnonminimal} \frac{\ddot{a}}{a}&=&-\frac{1}{6}[\hat{\rho}_{\phi }+\hat{\rho}_{\psi }+3(\hat{p}_{\phi }+\hat{p}_{\psi })],\\
\label{friedmanequationnonminimal} 3H^{2}&=&\hat{\rho}_{\phi }+\hat{\rho}_{\psi },
\end{eqnarray}
through the re-definition of the densities (\ref{rho_phi_minimal}), (\ref{rho_psi_minimal}) and pressures (\ref{p_phi_minimal}), (\ref{p_psi_minimal}) to
\begin{eqnarray}
\label{rho_phi_nonminimal} \hat {\rho}_{\phi } &=& \alpha ^{-1}\left( \frac{1}{2}\dot{\phi}^{2}+U(\phi
)+6\xi H\phi \dot{\phi}\right),\\
\label{rho_psi_nonminimal} \hat {\rho}_{\psi } &=& \alpha ^{-1}\left( \bar{\psi}(m+f\left(\phi\right))\psi +V(\bar{\psi}
,\psi )\right),\\
\label{p_phi_nonminimal} \hat{p}_{\phi } &=&\alpha ^{-1}\left( \frac{1}{2}\dot{\phi}^{2}-U(\phi
)-2\xi \left( \phi \ddot{\phi}+\dot{\phi}^{2}+2H\phi \dot{\phi}\right)
\right), \\
\label{p_psi_nonminimal} \hat{p}_{\psi } &=&\alpha ^{-1}\left(\frac{1}{2}\bar{\psi}\frac{\partial V%
}{\partial \bar{\psi}}+\frac{1}{2}\frac{\partial V}{\partial \psi }\psi -V(\bar{\psi},\psi )\right).
\end{eqnarray}

Note how the non-minimal parameter $\xi$ is more relevant for the equations related to the bosonic field (\ref{rho_phi_nonminimal}) and (\ref{p_phi_nonminimal}). Note also that we recover the minimal coupling equations (\ref{rho_phi_minimal} - \ref{p_psi_minimal}) when $\xi =0$ as it is expected.

On the other hand, we have the following equation obtained by adding (\ref{rho_phi_nonminimal}) and (\ref{p_phi_nonminimal}) and replacing (\ref{phisolution})
\begin{equation}\label{omega-phi}
\hat \rho _{\phi }\left( 1+\omega _{\phi }\right) =\alpha^{-1} \{ H^{2}\left( \phi ^{\prime2}a^{2}\left( 1-2\xi \right) -2\xi \phi \phi ^{\prime \prime }a^{2}\right)-\xi \phi \phi ^{\prime }a^{2}\frac{d}{da}H^{2}\},
\end{equation}
which will be used in numerical simulation to find the initial conditions for the scalar field $\phi$.

The equations of state are
\begin{eqnarray*}
\hat p_{\phi } &=&\omega _{\phi } \hat \rho _{\phi }, \\
\hat p_{\psi } &=&\omega _{\psi }\hat \rho _{\psi },
\end{eqnarray*}
which yields the following redefined conservation laws
\begin{eqnarray}
\label{dot_rho_phi_nonminimal} \dot{\hat{\rho}}_{\phi }+3H\left(\hat{\rho}_{\phi }+\hat{p}_{\phi
}\right) &=& -\frac{2\xi \phi \dot{\phi}}{1-\xi \phi ^{2}}\hat{\rho}_{\psi }-\hat{Q},\\
\label{dot_rho_psi_nonminimal} \dot{\hat{\rho}}_{\psi }+3H \left( \hat{\rho}_{\psi } + \hat{p}_{\psi} \right) &=& \frac{2\xi \phi \dot \phi}{1-\xi \phi ^{2}}\hat{\rho}_{\psi }+\hat{Q},
\end{eqnarray}
where $\hat{Q}=\alpha ^{-1}\bar{\psi}\frac{\partial f}{\partial \phi }\psi \dot{\phi}$ has been defined. In order to simplify calculations a dust type of equation for the fermionic field will be assumed, i. e. $\omega _{\psi }=0$ in equation (\ref{dot_rho_psi_nonminimal}).

One interesting point to observe is that, through these redefinitions the energy-momentum tensor is now conserved $\nabla ^{\mu }\hat{T}_{\mu \nu }=0$, with $\hat{T}_{\mu \nu}=\alpha ^{-1}T_{\mu \nu }$. At this point it is clear that we can interpret \textit{fermionic fields as sources of dark matter} and \textit{bosonic fields as sources of dark energy} as claimed above.

We see that positive acceleration imposes
\begin{equation}\label{accelerationcondition}
\hat{\rho}_{\phi }+\hat{\rho}_{\psi }+3\omega _{\phi }\hat{\rho}_{\phi }<0,
\end{equation}
and from this equation we find $\omega _{\phi }<-\frac{1}{3}\left( 1+r\right)$, in order to accelerated expansion to make sense, where we have defined $r =\frac{\hat{\rho}_{\psi }}{\hat{\rho}_{\phi }}$ called the coincidence parameter, with $r_{0}\approx\frac{3}{7}$ the actual value.

\section{Cosmological Solutions}

On the premise that minimal coupling can be achieve on non-minimal coupling equations imposing $\xi = 0$, we will focus on solving the general case, assuming an Ansatz for $\hat{Q}$. Let us consider the following cases found in the literature \cite{zimdahl,chimento,delcampo,cruz,lepe}
\begin{eqnarray}
\label{Qpsi} \hat{Q}=3\lambda H\hat{\rho}_{\psi },\\
\label{Qphi} \hat{Q}=3\lambda H\hat{\rho}_{\phi },
\end{eqnarray}
where $\lambda$ is a positive parameter, which means dark energy represented by the scalar field is transferring into dark matter fermions, as current observations suggest \cite{komatsu}.

Note that a dust type of model i.e. $\omega_{\psi} = 0$ or equivalently $p_{\psi} = 0$ imposed on the equation (\ref{p_psi_nonminimal}) implies
\begin{equation}\label{potencial1}
V\left(\bar \psi, \psi \right) = \frac{1}{2}\left(\bar \psi \frac{\partial V}{\partial \bar \psi} + \frac{\partial V}{\partial \psi} \psi \right),
\end{equation}
in accordance with equation (\ref{fermionscalarminimal}), let us suppose that the self-interaction fermionic potential is of the form  $V\left(\bar \psi,\psi \right)=V\left(\bar \psi \psi \right)$. This symmetry seems to be natural considering the need of the system to interact between matter and anti-matter in a way that is insensitive to a charge conjugation operation. This allows to write equation (\ref{potencial1}) in the following way
\begin{equation}\label{fermionicpotential}
V \left( \bar \psi \psi \right) = V \left( S \right) = b \, S,
\end{equation}
where $b$ is an integration constant.

On the other hand by replacing this result in equation (\ref{rho_psi_nonminimal}) and use equations (\ref{phisolution}) and (\ref{fermionicpotential}) it follows
\begin{equation}\label{constantecero}
V \left( S \right) = - m S - \beta c S,
\end{equation}
where we can identify $b = - m$ and $ c = 0 $. This is justified because the force derived from this potential must have a range of the order of $ \sim \frac{1}{m}$.

\begin{rmk}
In all the numerical solutions of the following subsections the parameters are fixed to $\omega_{\phi_0}=-0.98$, $\lambda=0.001$ and $\xi=\frac{1}{6}$ (conformal coupling). This is according to last astronomical observations \cite{komatsu}.
\end{rmk}

\subsection{Case I: $\hat{Q}=3\lambda H\hat{\rho}_{\psi }$}

Using (\ref{dot_rho_psi_nonminimal}) and (\ref{Qpsi}) an analytical solution for $\rho _{\psi }(a)$ is found
\begin{equation}\label{rhopsianalytic}
\rho _{\psi }=\rho _{\psi _{0}}a^{3\left( \lambda -1\right) },
\end{equation}
with
\begin{equation}\label{phianalytic}
\phi =\phi _{0}a^{3\lambda },
\end{equation}
where we have identified $\phi _{0}=\frac{\rho _{\psi _{0}}}{\beta S_{0}}$ and $a_{0}=1$. The $\lambda$ dependence arises as a direct consequence of the interaction term. Replacing these results on (\ref{dot_rho_phi_nonminimal}) we obtain an equation for $\rho _{\phi }$ that can be solved numerically.

From equation (\ref{omega-phi}) an initial condition for $\phi $ in terms of observational and the model parameters $\xi, \lambda$ is found to be
\begin{equation}\label{phioIC}
\phi _{0}=\pm \sqrt{\frac{\left( 1+\omega _{0}\right) \frac{\hat{\rho}_{\phi_{0}}}{\hat{\rho}_{c}}}{\left( 3\lambda ^{2}-12\xi \lambda ^{2}+2\xi \lambda\right) +3\xi \lambda \left( 1+\omega _{0}\frac{\hat{\rho}_{\phi _{0}}}{\hat{
\rho}_{c}}\right) +\xi \left( 1+\omega _{0}\right) \frac{\hat{\rho}_{\phi_{0}}}{\hat{\rho}_{c}}}},
\end{equation}
where $\hat{\rho}_{c}=3H_{0}^{2}$ is the actual density critical value at $t=t_{0}$. We
note also that in this case the condition $1-\xi \phi ^{2}=0$ produces $a_{c}=\left( \xi \phi _{0}^{2}\right) ^{-\frac{1}{6\lambda }}$, it follows
\begin{eqnarray}
\nonumber \phi \left( a_{c} \right) &=& \phi_{c} \\
\nonumber &=& \pm \sqrt{\frac{1}{\xi}}\\
\label{phicritico} &=& \pm \sqrt{6}.
\end{eqnarray}
Case I presents $\phi_{0} = 2.22 < \phi_{c}$.

\begin{figure}[h]
  \centering
  \subfloat[] {\label{densidadI}\framebox{\includegraphics[width=0.4\textwidth]{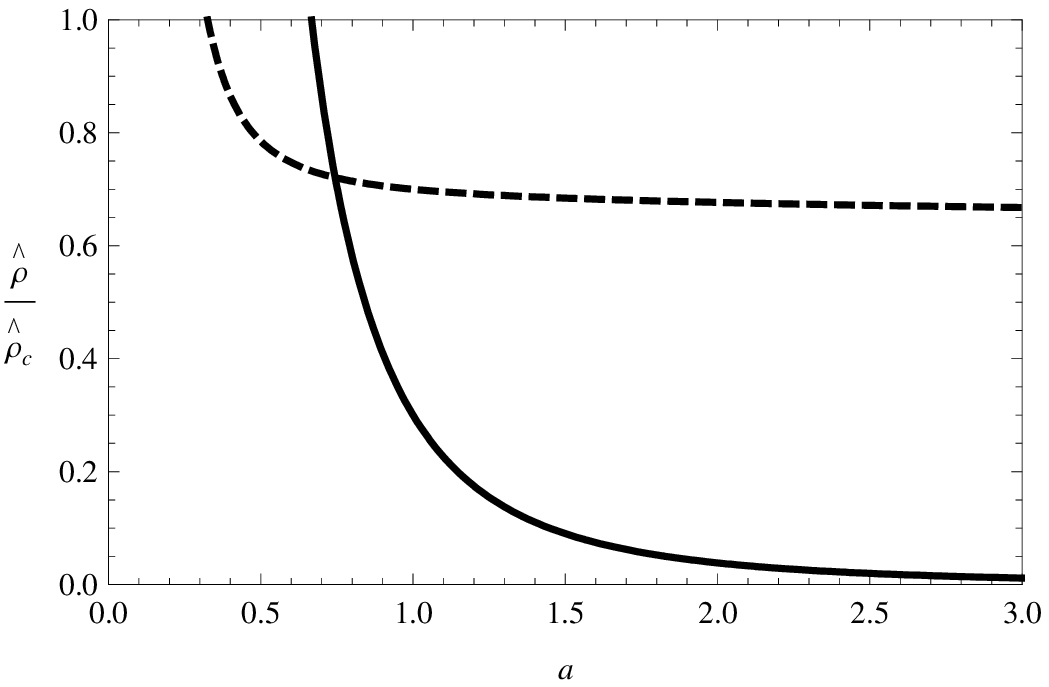}}}
  \quad\quad\quad
  \subfloat[]{\label{H}\framebox{\includegraphics[width=0.4\textwidth]{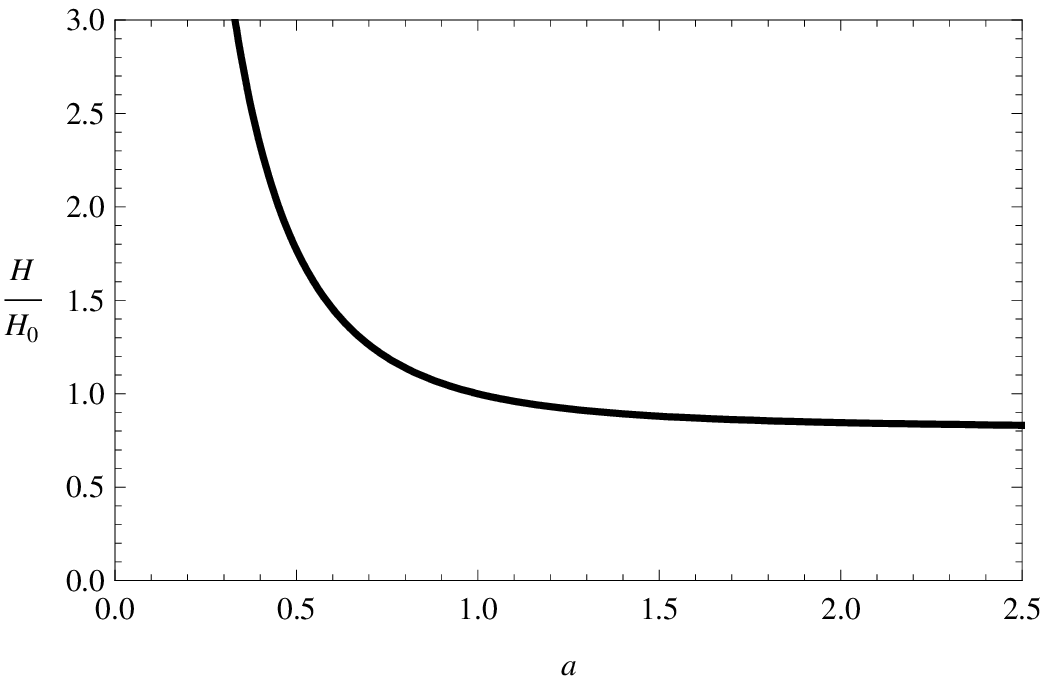}}}
  \caption{\label{densidadIyH}}
\end{figure}

\begin{figure}[h]
  \centering
  \subfloat[]{\label{aceleracio}\framebox{\includegraphics[width=0.4\textwidth]{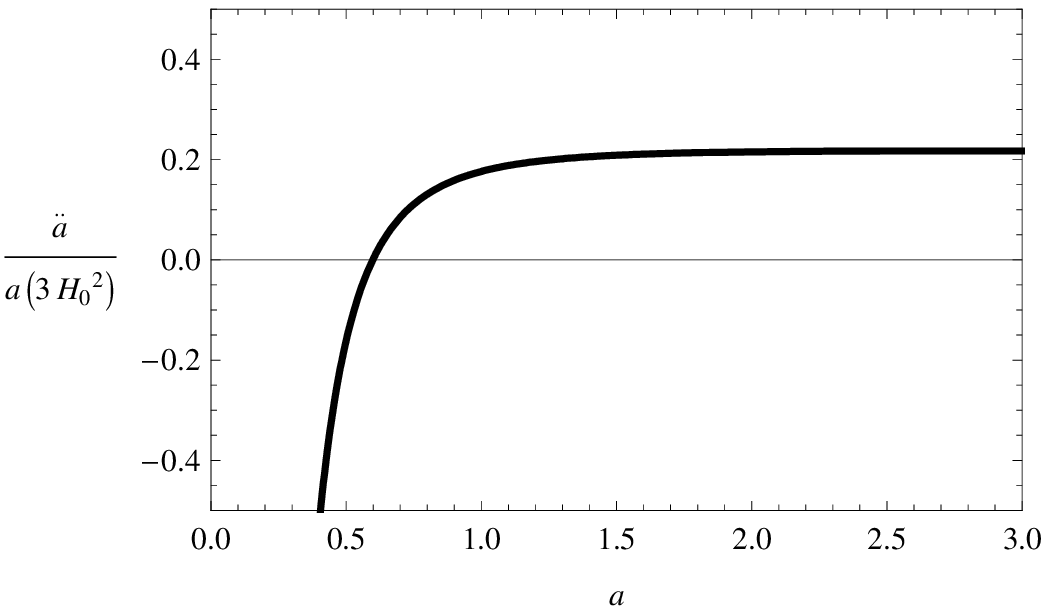}}}
  \quad\quad\quad
  \subfloat[]{\label{omegaI}\framebox{\includegraphics[width=0.4\textwidth]{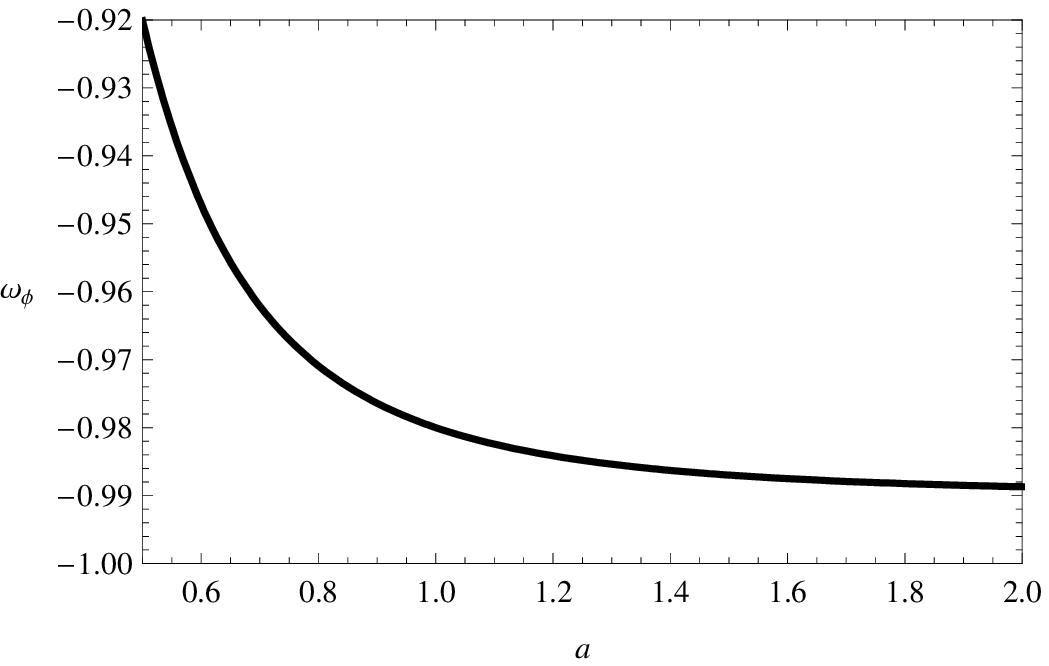}}}
  \caption{\label{aceleracioyomegaI}}
\end{figure}

As can be seen, Figure \ref{densidadI} shows the redefined densities $\hat{\rho}_\phi$ (dashed line) and $\hat{\rho}_\psi$ (solid line) behave in a way that in the distant future only dark energy survives, i. e. scalar field. Figure \ref{H} shows a de Sitter like universe evolution in a distant future. A transition from decelerated expansion to accelerated expansion occurs at $a \approx 0.6$, which is shown in Figure \ref{aceleracio}, to go along an approximately constant value of  $\frac{\ddot{a}}{a3H_0^2}= 0.2$.

\begin{figure}[h]
  \centering
  \subfloat[]{\label{r}\framebox{\includegraphics[width=0.38\textwidth]{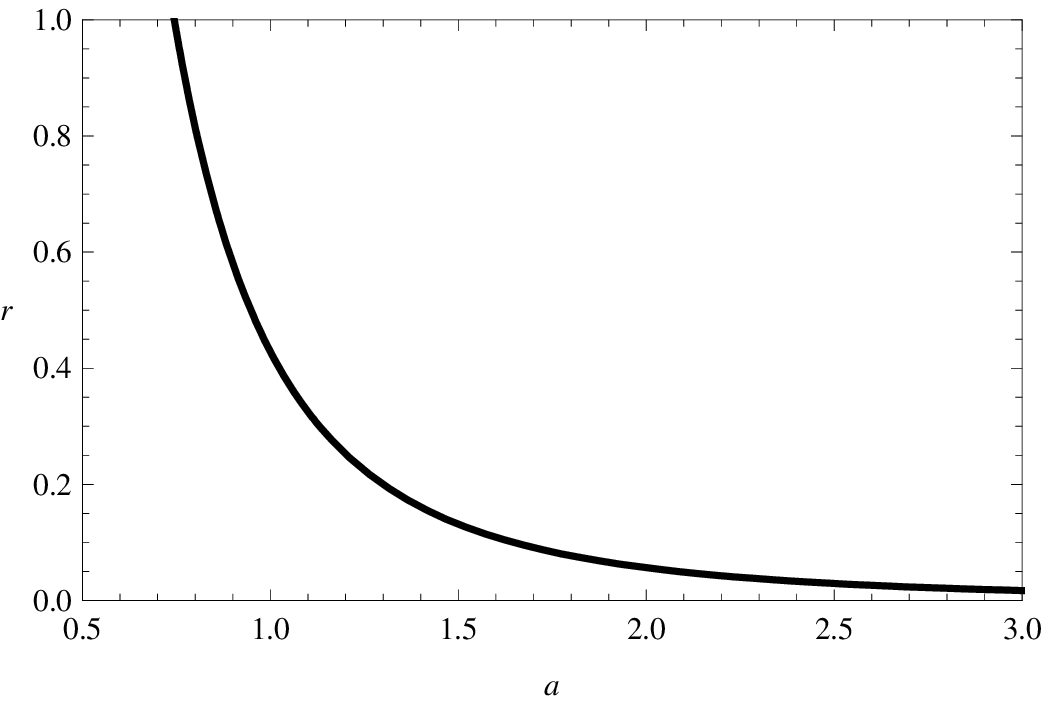}}}
  \quad\quad\quad
  \subfloat[]{\label{Q}\framebox{\includegraphics[width=0.38\textwidth]{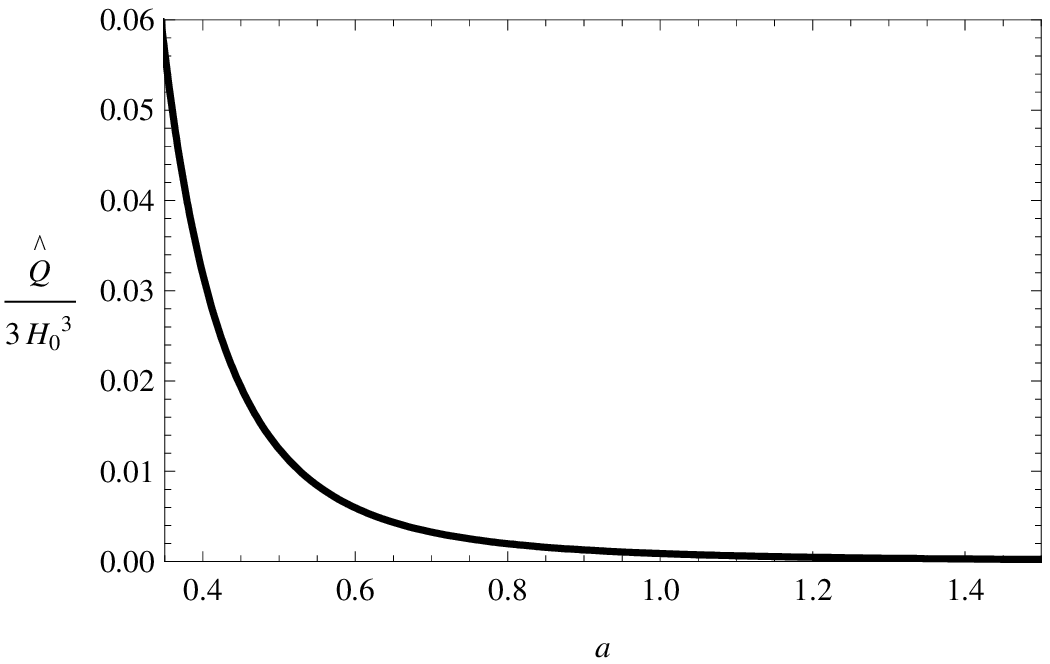}}}
  \caption{\label{ryQ}}
\end{figure}

Under the conditions of the simulation, in Figure \ref{omegaI} a quintessence type of solution is obtained for $\omega > -1$ which is according to experimental facts. However, using $\omega_0 < -1 $ a phantom type of behavior is obtained which has no physical foundation. Figure \ref{r} shows a decreasing $r$, which is expected for a compete dark energy dominance in the future.

\begin{figure}[h]
  \centering
  \subfloat[]{\label{phiI}\framebox{\includegraphics[width=0.38\textwidth]{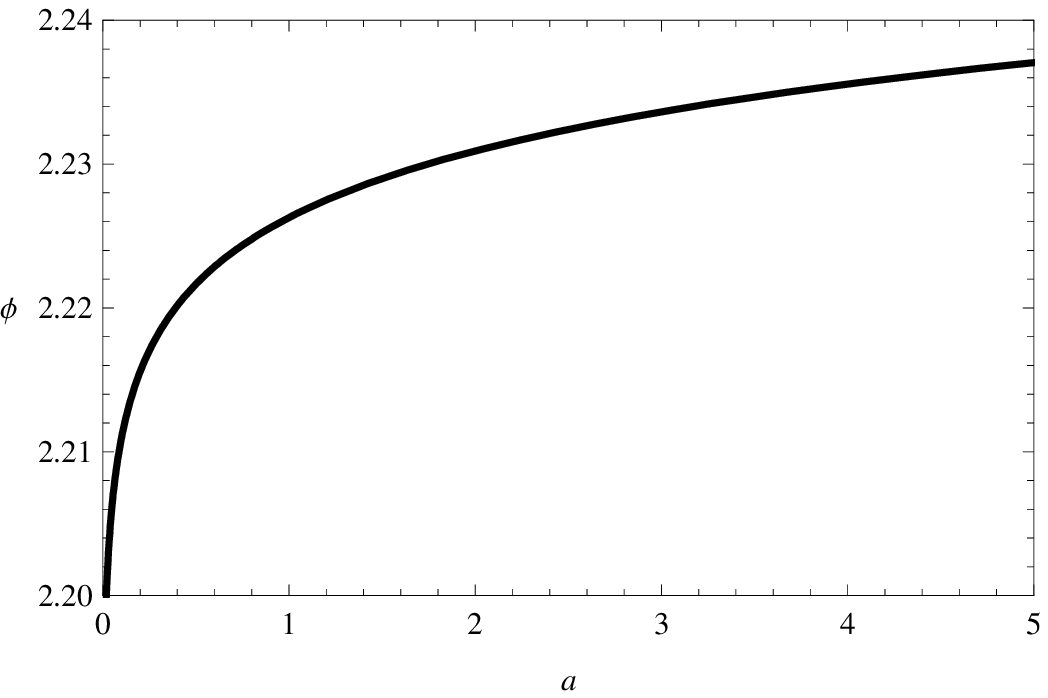}}}
  \quad\quad\quad
  \subfloat[]{\label{ptencial}\framebox{\includegraphics[width=0.38\textwidth]{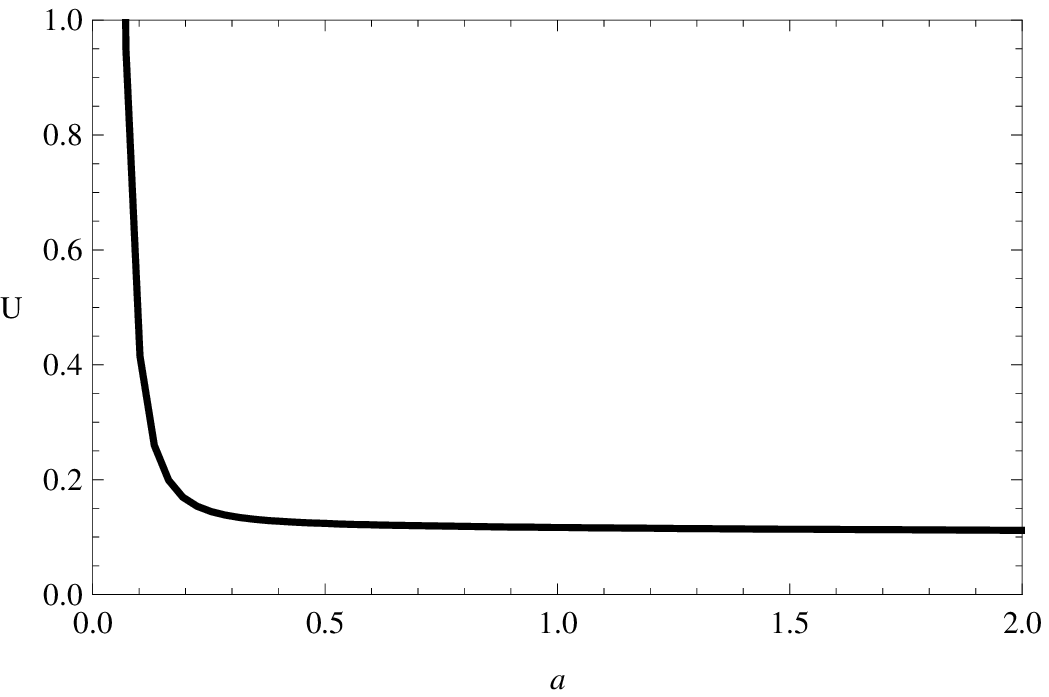}}}
  \caption{\label{phiIyptencial}}
\end{figure}

Figure \ref{Q} indicates a decreasing rate of change from dark matter to dark energy, to the point in which dark matter practically vanishes in the future when the source has totally transferred its energy. Figure \ref{phiI} shows that $\phi$ grows rapidly for $a \ll 1$ growing then to a lesser rate of change than initially observed, according to the behavior of the densities $\hat \rho_\phi$ and $\hat \rho_\psi$, $\hat Q$  and the coincidence parameter $r$. In Figure \ref{ptencial} the potential $V\left( a \right)$ is constructed, the self-interaction is almost null for $a \gg 1$.

\subsection{Case II: $\hat{Q}=3\lambda H\hat{\rho}_{\phi }$}

In this case the solutions will be obtained numerically due to the nonlinearity of the system of differential equations. Similarly as for the Case I from (\ref{omega-phi}) the initial condition for $\phi$ is obtained. Case II presents two branches for possible initial conditions for $\phi_0$
\begin{eqnarray*}
\phi_0 &=& 3.45 > \phi_c, \\
\phi_0 &=& 2.43 < \phi_c.
\end{eqnarray*}

The first branch has a very similar behavior to Case I and is a quintessence universe. The second alternative is more keen to study due to the fact that it has a phantom type of behavior and for parameters of the model allows the crossing to quintessence.

\begin{figure}[h]
  \centering
  \subfloat[]{\label{densidadesII}\framebox{\includegraphics[width=0.38\textwidth]{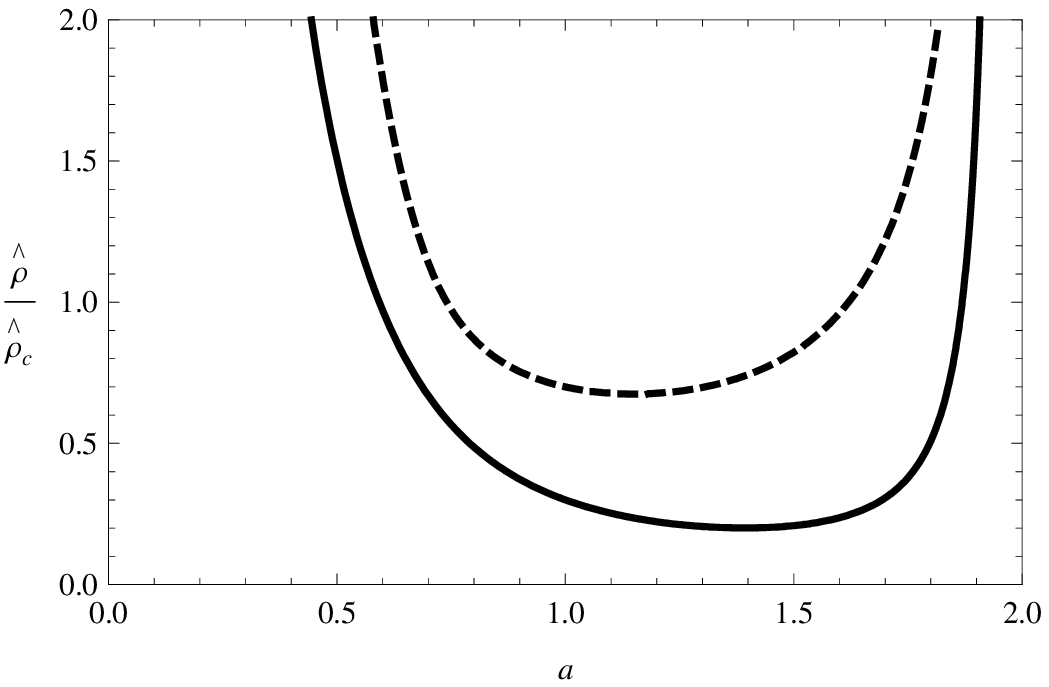}}}
  \quad\quad\quad
  \subfloat[]{\label{hII}\framebox{\includegraphics[width=0.38\textwidth]{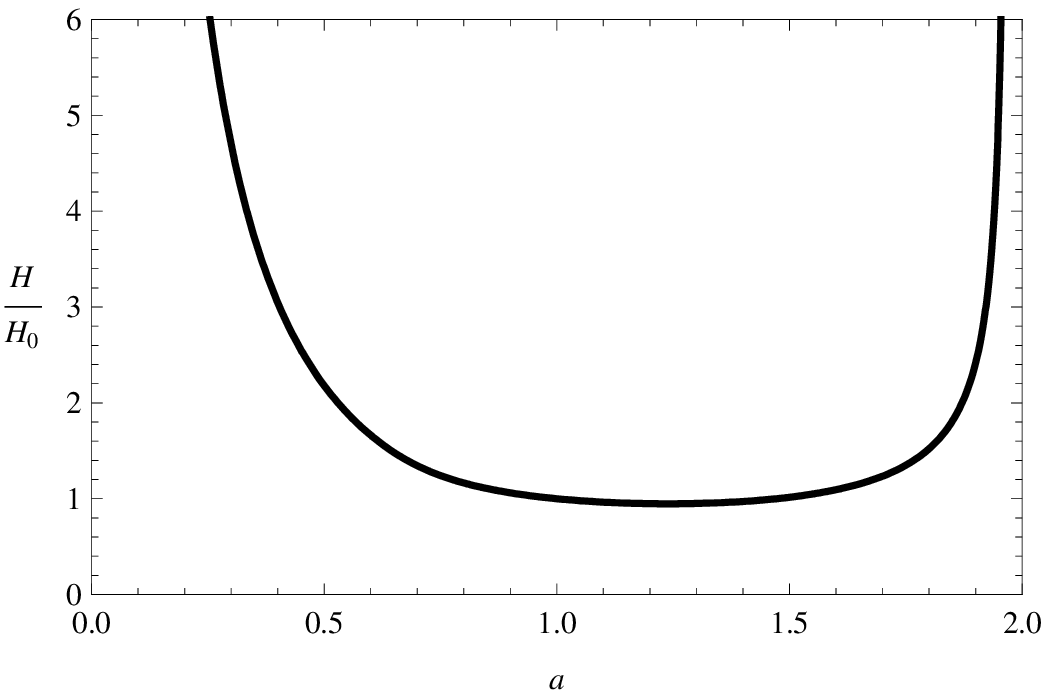}}}
  \caption{\label{densidadesIIyhII}}
\end{figure}

In Figure \ref{densidadesII} the redefined densities $\hat \rho_\phi$ and $\hat \rho_\psi$ show a good behavior up to the neighborhood of the critical point $a_c$ where both solutions diverge. The overall tendency is that dark energy density is always bigger than the matter density. Recall that both solutions are valid only in the range of the second branch. In Figure \ref{hII} it is clearly seen that there is a divergence due to the factor $\alpha^{-1}$ in the solutions.

\begin{figure}[h]
  \centering
  \subfloat[]{\label{aceleracionII}\framebox{\includegraphics[width=0.38\textwidth]{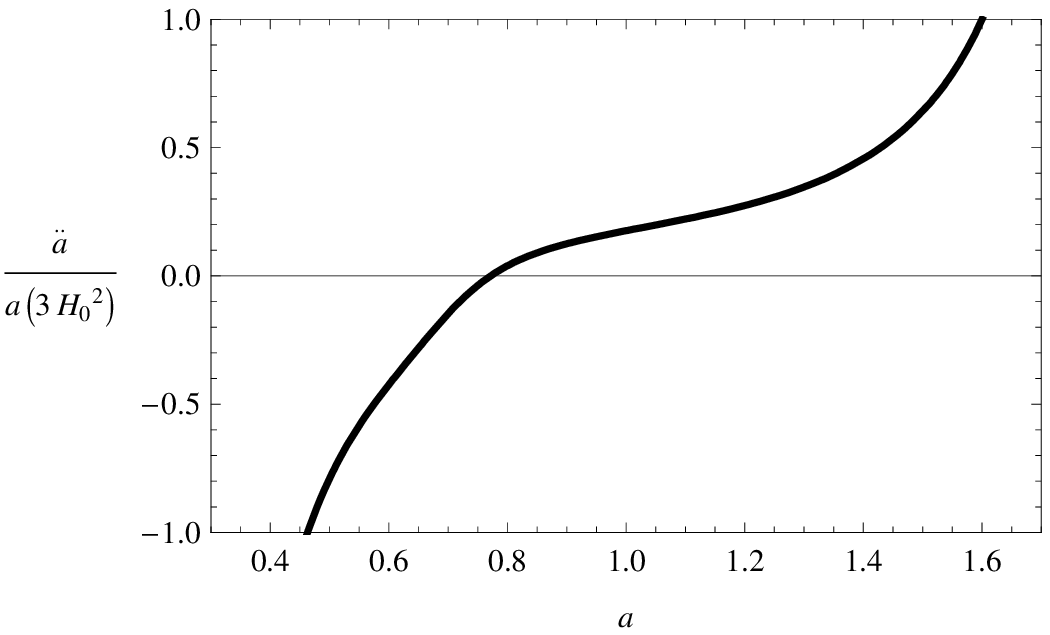}}}
  \quad\quad\quad
  \subfloat[]{\label{omegaII}\framebox{\includegraphics[width=0.38\textwidth]{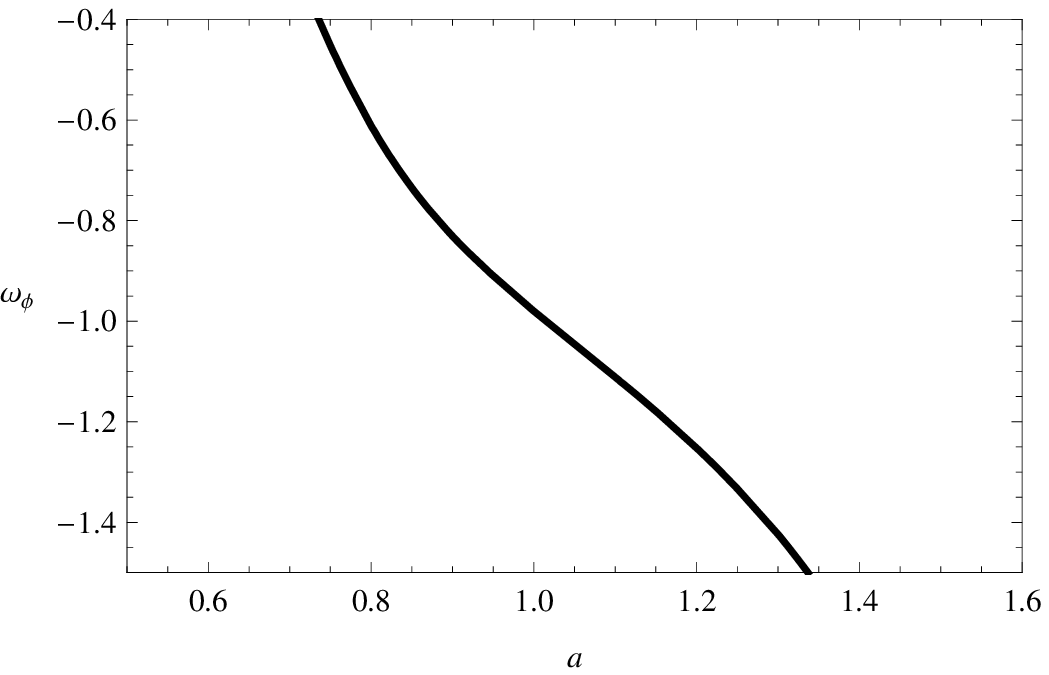}}}
  \caption{\label{aceleracionIIyomegaII}}
\end{figure}

Figure \ref{aceleracionII} shows that this solution in the remote past the universe was decelerated and then at $a\approx0.7$ begins to accelerate with a great increase of its growing rate in the neighborhood of $a_c$. Adding a non-minimal coupling favors the phantom evolution allowing the crossing of the barrier $\omega=-1$, which is clearly shown in Figure \ref{omegaII}, as claim the recent observations \cite{komatsu}.

\begin{figure}[h]
  \centering
  \subfloat[]{\label{rII}\framebox{\includegraphics[width=0.38\textwidth]{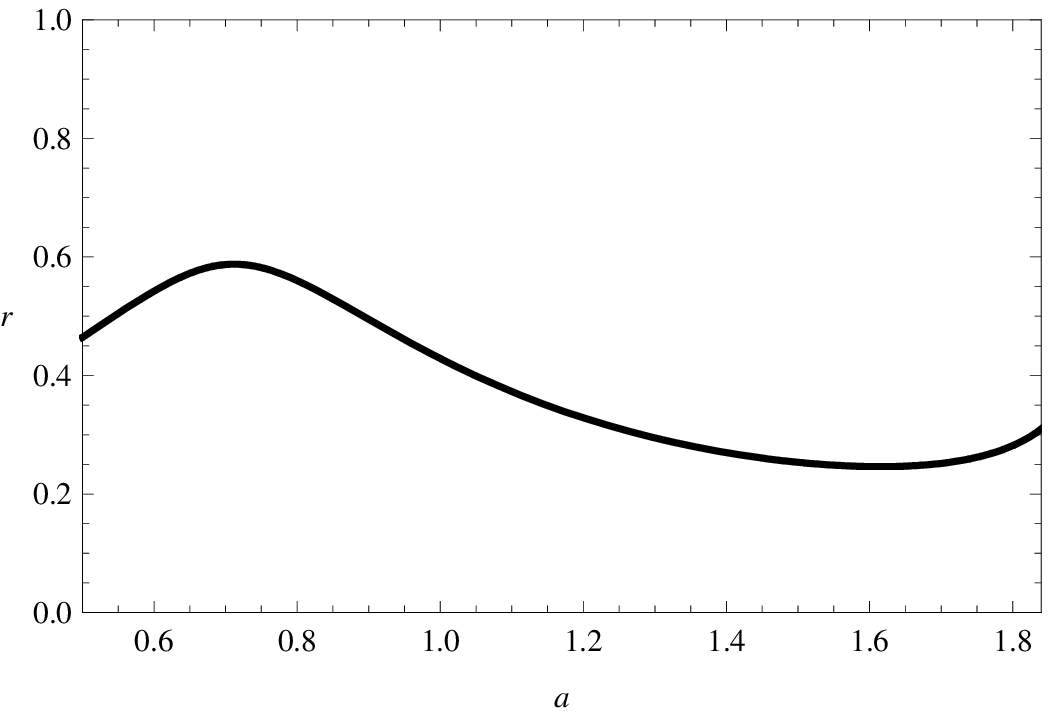}}}
  \quad\quad\quad
  \subfloat[]{\label{QII}\framebox{\includegraphics[width=0.38\textwidth]{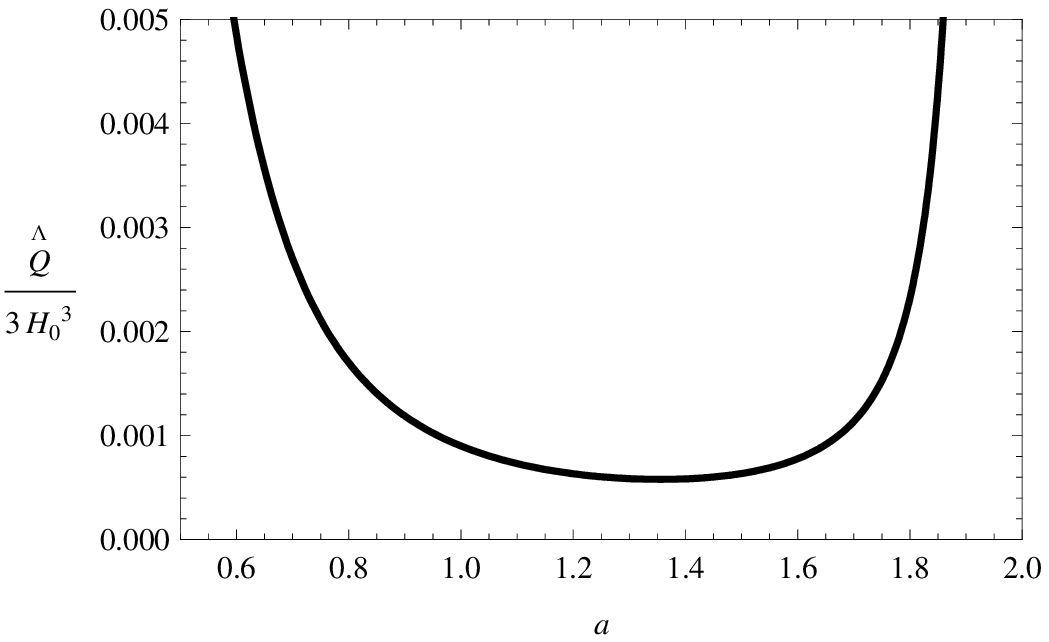}}}
  \caption{\label{rIIyQII}}
\end{figure}

Figure \ref{rII} exhibits similar behavior to those of the previous solutions due to the presence of a singularity at $\phi_c$, however in the observational range there is a tendency to stabilization and constancy of the solution. Note also that the maximum is observed to be at the same turning point of the acceleration $a \approx 0.7$. Figure \ref{QII} presents the same divergent behavior in the neighborhood of $\phi_c$ and has a tendency to remain constant before the singularity.

\begin{figure}[h]
  \centering
  \subfloat[]{\label{phiII}\framebox{\includegraphics[width=0.38\textwidth]{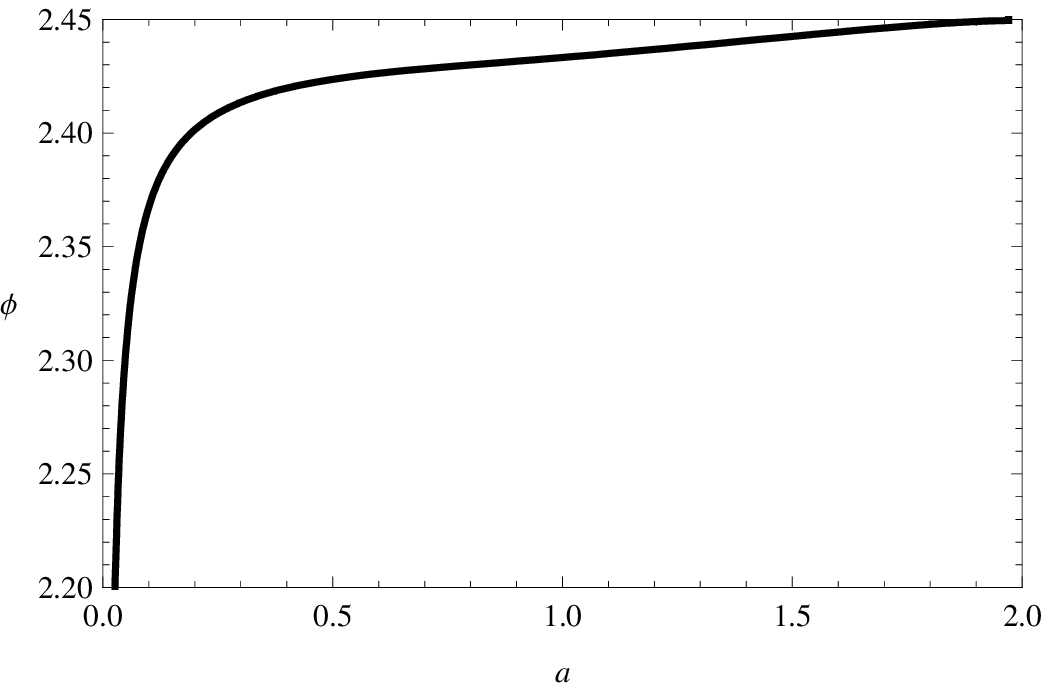}}}
  \quad\quad\quad
  \subfloat[]{\label{potencialII}\framebox{\includegraphics[width=0.38\textwidth]{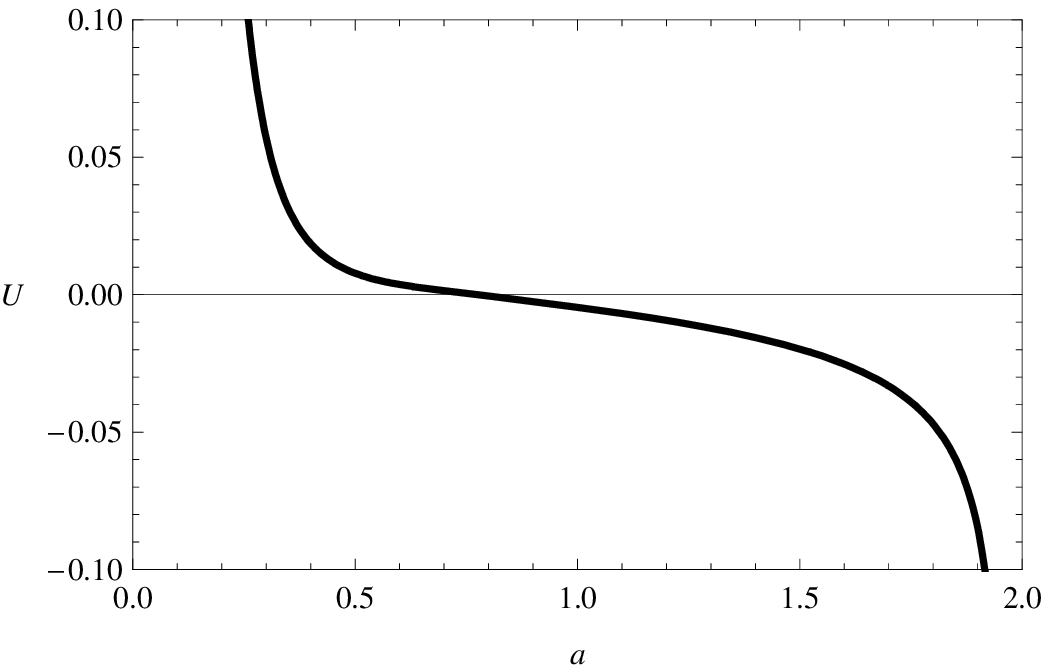}}}
  \caption{\label{phiIIypotencialII}}
\end{figure}

Figure \ref{phiII} shows that under general lines the behavior of the scalar field is the same as that in Case I. The self interaction potential in Figure \ref{potencialII} shows a dual behavior with respect of the expansion parameter in such a way that it is attractive for $a < 0.7$ and repulsive for $a > 0.7$. It is worth to mention that the transition happens just in the transition from decelerated to accelerated universe.

\section{Discussion}

\begin{figure}[htp]
\centering
  \framebox{\includegraphics[width=4.5in,height=3in]{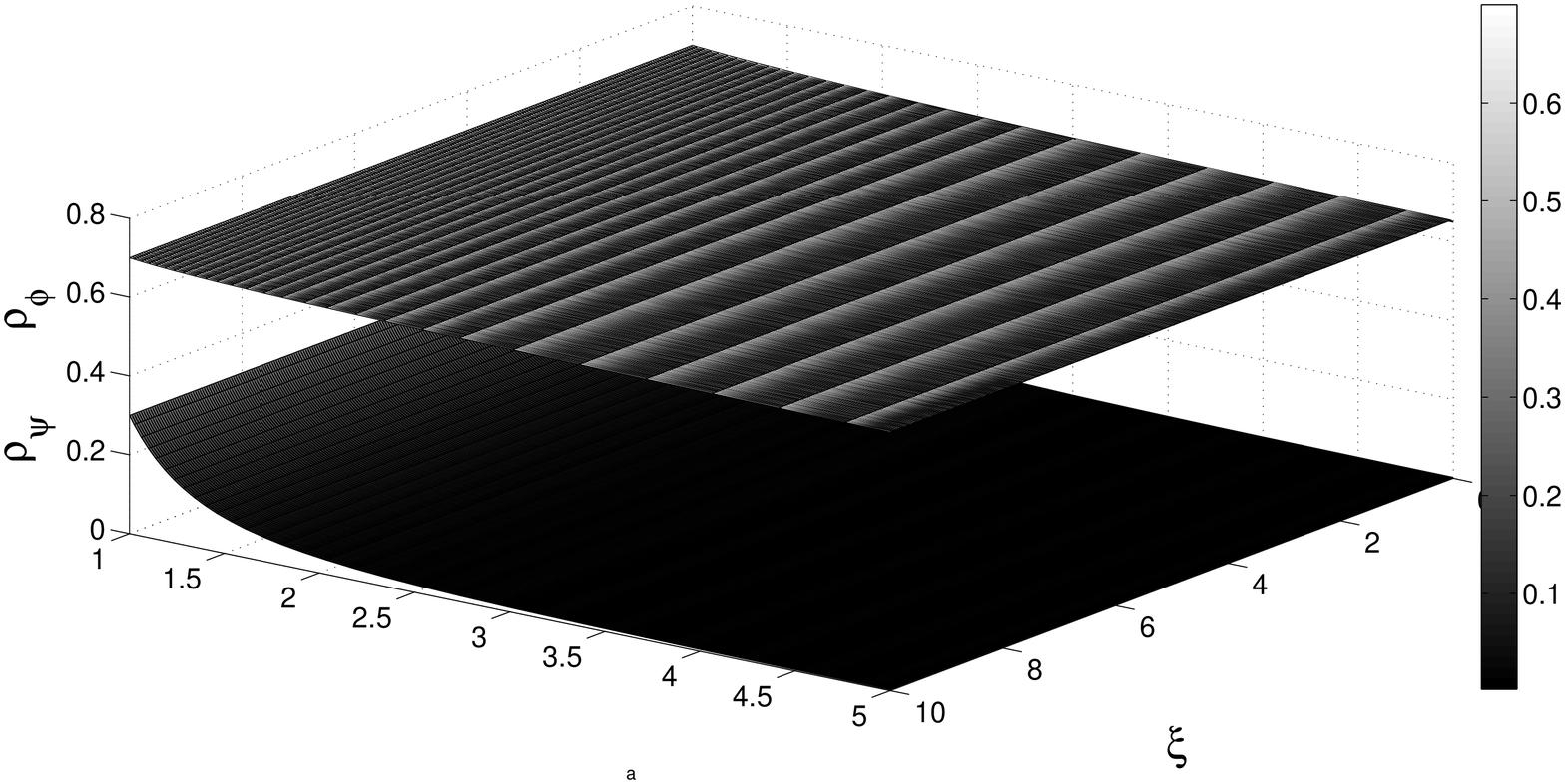}}\\
  \caption{\label{densidadesvsavsxi}}
\end{figure}

As it can be seen in Case I, minimal coupling does not allow to pass over the phantom barrier $\omega=-1$ for the initial conditions chosen. This seems to be the general behavior, reinforced by the exploration of the parameter space for large ranges. As far as numerical simulations permit, the behavior for this type of model in the future is quintessence. In Case II, non-minimal coupling allows to pass over the $\omega=-1$ barrier for certain initial conditions, although the model seems to be very insensitive of the value of the coupling parameter $\xi$ (see  Figure \ref{densidadesvsavsxi}) . This type of model delivers a type III cosmological singularity produced mainly because of the $\alpha$ factor. Both solutions found are according to the observational measures.

The incorporation of the scalar field as source of dark energy and the fermionic field as source of dark matter seems to model very adequately the dynamics as well as the interaction in the cosmological models found.

Having said that, in Case II, the $\xi$ parameter does not take such a prominent effect on the energy density solutions. These behavior is shown in Figure \ref{densidadesvsavsxi}, where the dependence of the redefined densities to the cosmological parameter $a$ and the model parameter $\xi$ are plotted. This $\xi$ parameter insensibility can be take under account to find an almost-analytical solution in the case of cosmological constant universe. In the appendix is shown how to find a Green function that leads to the following almost-analytical solution for the scalar field
\begin{equation}\label{appendixsolution}
\phi_{\xi} \left( t \right) = \sqrt {\frac{{{\theta _\xi }}}{{2\pi \xi }}} {\left[ {\int\limits_{0}^t {\frac{{dt'}}{{a\left( {t'} \right)}}} } \right]^{\frac{1}{2}}},
\end{equation}
where
\begin{equation}\label{theta_xi}
{\theta _\xi } = \sum\limits_{i = 1}^\infty {{a_i}{\xi ^i}}.
\end{equation}

The importance of this solution is that being valid in the barrier $\omega_\phi = -1$ permits the comparison of solutions on either side of the barrier, this let us propose the following remark.

\begin{rmk}
Let the solutions of a cosmological model derived from the Lagrangian (\ref{lagrangiana}) be valid in the interval $[a,b] \subseteq \mathbb{R}$, then a scalar field solution fulfilling
\begin{equation}\label{condition1}
\phi \left( t \right) > \phi_\xi \left( t \right)\;\;\;\forall\;t \in [a,b],
\end{equation}
belongs to a phantom type of universe, whereas a scalar field solution fulfilling
\begin{equation}\label{condition2}
\phi \left( t \right) < \phi_\xi \left( t \right)\;\;\;\forall\;t \in [a,b],
\end{equation}
belongs to a quintessence type of universe
\end{rmk}

To complete the view, an almost-analytical solution is obtained for the Hubble parameter in terms of the scalar function
\begin{equation}\label{Hcosmolicalconstant}
\nonumber {H} = \frac{\beta {S_0}}{2}\int\limits_{t_0}^t {\frac{{{\phi _\xi }\left( t' \right)}}{{\left( {\xi \phi _\xi ^2\left( t' \right) - 1} \right){a'\left( t \right)^3}}}dt'},
\end{equation}
see appendix for details.

\section{Appendix}

We will show here how to find a general Green's function for a non-minimal coupling at the boundary of cosmological constant. Using equations (\ref{rho_phi_nonminimal}), (\ref{p_phi_nonminimal}), and the equations of state we have:
\begin{equation}\label{rhowithoperator}
{\rho _\phi }\left( {1 + {\omega _\phi }} \right) = \left( {{{\dot \phi }^2} - 2\xi {L_H}\left( {{\textstyle{1 \over 2}}{\phi ^2}} \right)} \right),
\end{equation}
where the operator ${L_H} = \frac{{{d^2}}}{{d{t^2}}} + H\left( t \right)\frac{d}{{dt}}$ has been defined.

Note that on the left hand of equation (\ref{rhowithoperator}), as $\rho_\phi$ is positive defined then for a cosmological constant solution ($\omega_\phi=-1$) follows
\begin{equation}\label{cosmologicalconstantdiffeq}
\frac{1}{{2\xi }}{{\dot \phi }_\xi^2}\, = {L_H}\left( {{\textstyle{1 \over 2}}{\phi_{\xi} ^2}} \right).
\end{equation}

The formal Green's function solution for the previous equation is
\begin{equation}\label{Greenintegral}
{\phi_{\xi} ^2} = \frac{1}{\xi }\int\limits_0^\infty  {G\left( {t,s} \right){{\left( {\frac{{d\phi_{\xi}}}{{ds}}} \right)}^2}ds},
\end{equation}
where the kernel $G:R \bigotimes R \rightarrow R$ fulfills ${\mathfrak{L}_H}G\left( {t,s} \right) = \delta \left( {t - s} \right)$ and ${\mathfrak{L}_H} = \frac{{{\partial ^2}}}{{\partial {t^2}}} + H\left( t \right)\frac{\partial }{{\partial t}}$.

Let us define the following auxiliary function $K:R \bigotimes R \rightarrow R$: $K\left( {t,s} \right) = \frac{\partial }{{\partial t}}G\left( {t,s} \right)$ which fulfills the following differential equation
\begin{equation}\label{auxiliarykerneldifferentialequation}
\frac{\partial }{{\partial t}}K\left( {t,s} \right) + H\left( t \right)K\left( {t,s} \right) = \delta \left( {t - s} \right),
\end{equation}
by performing a Fourier Transform equation (\ref{auxiliarykerneldifferentialequation}) can be written
\begin{equation}\label{auxiliarykernelFT}
\frac{\partial }{{\partial t}}K\left( {t,x} \right) + H\left( t \right)K\left( {t,x} \right) = \frac{1}{{\sqrt {2\pi } }}\exp \left( { - ixt} \right).
\end{equation}

The equation (\ref{auxiliarykernelFT}) is the Bernoulli's differential equation, which has a known solution
\begin{equation}\label{bernoullisolution}
K\left( {t',x} \right) = \frac{1}{{\sqrt {2\pi } }}\frac{{\int\limits_{0}^{t'} {\exp \left( {\int\limits_{0}^{t'''} {H\left( {t'''} \right)dt'''} } \right)\exp \left( { - it''x} \right)dt''} }}{{\exp \left( {\int\limits_{0}^{t'} {H\left( {t''} \right)dt''} } \right)}},
\end{equation}
by using the Hubble's parameter and an inverse Fourier Transform it follows
\begin{equation}\label{kernelderivative}
K\left( {t',s} \right) = \frac{1}{{{2\pi } a\left( {t'} \right)}}\int\limits_{0}^{t'} {a\left( {t''} \right)\delta \left( {t'' - s} \right)dt''}  = \frac{1}{{{2\pi } }}\frac{{a\left( s \right)}}{{a\left( {t'} \right)}},
\end{equation}
then the kernel is finally found to be
\begin{equation}\label{finalgreenfunction}
G\left( {t,s} \right) = \frac{{a\left( s \right)}}{{{2\pi } }}\int\limits_{0}^t {\frac{{dt'}}{{a\left( {t'} \right)}}}.
\end{equation}

By replacing (\ref{finalgreenfunction}) on (\ref{Greenintegral})
\begin{equation}
\nonumber {\phi_{\xi} ^2}\left( t \right) = \frac{1}{{2\pi \xi }}\int\limits_{0}^t {\frac{{dt'}}{{a\left( {t'} \right)}}} \left[ {\int\limits_{0}^\infty  {a\left( s \right){{\left( {\frac{{d\phi_{\xi} }}{{ds}}} \right)}^2}ds} } \right],
\end{equation}
note that the expression in brackets only depends on the parameter $\xi$ which allows to write
\begin{equation}
\nonumber \phi_{\xi} \left( t \right) = \sqrt {\frac{{{\theta _\xi }}}{{2\pi \xi }}} {\left[ {\int\limits_{0}^t {\frac{{dt'}}{{a\left( {t'} \right)}}} } \right]^{\frac{1}{2}}},
\end{equation}
which is equation (\ref{appendixsolution}), where ${\theta _\xi } = \int\limits_{0}^\infty  {a\left( s \right){{\left( {\frac{{d{\phi _\xi }}}{{ds}}} \right)}^2}ds}$ has been defined.

Assuming ${\theta _\xi }$ is analytic then it must be of first order in $\xi$ or higher to ensure convergence, as is readily seen on equation (\ref{rhowithoperator}). A Taylor expansion is therefore of the form:
\begin{equation}
\nonumber {\theta _\xi } = \sum\limits_{i = 1}^\infty {{a_i}{\xi ^i}},
\end{equation}
which is the equation (\ref{theta_xi}).

On the other hand, equation (\ref{omega-phi}) can be written
\begin{equation}
\nonumber {\left( {{\rho _\phi } + {\rho _\psi }} \right)^\prime } + 3\frac{{{\rho _\psi }}}{a} + \frac{\xi }{{\left( {1 - \xi \phi _\xi ^2\left( a \right)} \right)}}\left( {{\rho _\phi } + {\rho _\psi }} \right)\frac{d}{{da}}\left( {\phi _\xi ^2\left( a \right)} \right) = 0,
\end{equation}
which by aim of equation (\ref{phisolution}) takes the form
\begin{equation}
\nonumber 6HH'\left( {1 - \xi \phi _\xi ^2\left( a \right)} \right) + 3\frac{{\beta {S_0}}}{{{a^4}}}{\phi _\xi }\left( a \right) = 0,
\end{equation}
which finally can be written
\begin{equation}
\nonumber {H} = \frac{\beta {S_0}}{2}\int\limits_{t_0}^t {\frac{{{\phi _\xi }\left( t' \right)}}{{\left( {\xi \phi _\xi ^2\left( t' \right) - 1} \right){a'\left( t \right)^3}}}dt'},
\end{equation}
which is equation (\ref{Hcosmolicalconstant}).

\section{Acknowledgements}

This research was supported by DIR01.11 037.334/2011 PUCV and Fondecyt 1110076 (SL), also DI10-0009 of Direcci\'on de Investigaci\'on y Desarrollo, Universidad de La Frontera (FP), and also by DI11-0071 (YV).

\section*{References}
\bibliographystyle{elsarticle-num}
\biboptions{comma,square}
\bibliography{./bosonfermionblilbio}

\begin{thebibliography}{10}
\expandafter\ifx\csname url\endcsname\relax
  \def\url#1{\texttt{#1}}\fi
\expandafter\ifx\csname urlprefix\endcsname\relax\def\urlprefix{URL }\fi
\expandafter\ifx\csname href\endcsname\relax
  \def\href#1#2{#2} \def\path#1{#1}\fi

\bibitem{komatsu}
E.~Komatsu, et~al., {Seven-Year Wilkinson Microwave Anisotropy Probe (WMAP)
  Observations: Cosmological Interpretation}, Astrophys. J. Suppl. 192 (2011)
  18.
\newblock \href {http://arxiv.org/abs/1001.4538} {\path{arXiv:1001.4538}},
  \href {http://dx.doi.org/10.1088/0067-0049/192/2/18}
  {\path{doi:10.1088/0067-0049/192/2/18}}.

\bibitem{Buchbinder}
I.~L. Buchbinder, S.~D. Odintsov, I.~L. Shapiro, {Effective action in quantum
  gravity}Bristol, UK: IOP (1992) 413 p.

\bibitem{Elizalde}
E.~Elizalde, S.~D. Odintsov, {Renormalization group improved effective
  potential for finite grand unified theories in curved space-time}, Phys.
  Lett. B333 (1994) 331--336.
\newblock \href {http://arxiv.org/abs/hep-th/9403132}
  {\path{arXiv:hep-th/9403132}}, \href
  {http://dx.doi.org/10.1016/0370-2693(94)90151-1}
  {\path{doi:10.1016/0370-2693(94)90151-1}}.

\bibitem{Muta}
T.~Muta, S.~D. Odintsov, {Model dependence of the nonminimal scalar graviton
  effective coupling constant in curved space-time}, Mod. Phys. Lett. A6 (1991)
  3641--3646.
\newblock \href {http://dx.doi.org/10.1142/S0217732391004206}
  {\path{doi:10.1142/S0217732391004206}}.

\bibitem{faraoni}
V.~Faraoni, {Inflation and quintessence with nonminimal coupling}, Phys. Rev.
  D62 (2000) 023504.
\newblock \href {http://arxiv.org/abs/gr-qc/0002091}
  {\path{arXiv:gr-qc/0002091}}, \href
  {http://dx.doi.org/10.1103/PhysRevD.62.023504}
  {\path{doi:10.1103/PhysRevD.62.023504}}.

\bibitem{nojiri}
S.~Nojiri, S.~D. Odintsov, {Inhomogeneous equation of state of the universe:
  Phantom era, future singularity and crossing the phantom barrier}, Phys. Rev.
  D72 (2005) 023003.
\newblock \href {http://arxiv.org/abs/hep-th/0505215}
  {\path{arXiv:hep-th/0505215}}, \href
  {http://dx.doi.org/10.1103/PhysRevD.72.023003}
  {\path{doi:10.1103/PhysRevD.72.023003}}.

\bibitem{delamacorra}
A.~de~la Macorra, {Interacting dark energy: Decay into fermions}, Astropart.
  Phys. 28 (2007) 196--204.
\newblock \href {http://arxiv.org/abs/astro-ph/0702239}
  {\path{arXiv:astro-ph/0702239}}, \href
  {http://dx.doi.org/10.1016/j.astropartphys.2007.05.005}
  {\path{doi:10.1016/j.astropartphys.2007.05.005}}.

\bibitem{micheletti}
S.~Micheletti, E.~Abdalla, B.~Wang, {A Field Theory Model for Dark Matter and
  Dark Energy in Interaction}, Phys. Rev. D79 (2009) 123506.
\newblock \href {http://arxiv.org/abs/0902.0318} {\path{arXiv:0902.0318}},
  \href {http://dx.doi.org/10.1103/PhysRevD.79.123506}
  {\path{doi:10.1103/PhysRevD.79.123506}}.

\bibitem{ribas}
M.~O. Ribas, F.~P. Devecchi, G.~M. Kremer, {Fermions as sources of accelerated
  regimes in cosmology}, Phys. Rev. D72 (2005) 123502.
\newblock \href {http://arxiv.org/abs/gr-qc/0511099}
  {\path{arXiv:gr-qc/0511099}}, \href
  {http://dx.doi.org/10.1103/PhysRevD.72.123502}
  {\path{doi:10.1103/PhysRevD.72.123502}}.

\bibitem{nozari}
K.~Nozari, S.~D. Sadatian, {Late-time acceleration and Phantom Divide Line
  Crossing with Non-minimal Coupling and Lorentz Invariance Violation}, Eur.
  Phys. J. C58 (2008) 499--510.
\newblock \href {http://arxiv.org/abs/0809.4744} {\path{arXiv:0809.4744}},
  \href {http://dx.doi.org/10.1140/epjc/s10052-008-0767-3}
  {\path{doi:10.1140/epjc/s10052-008-0767-3}}.

\bibitem{zimdahl}
W.~Zimdahl, D.~Pavon, {Interacting quintessence}, Phys. Lett. B521 (2001)
  133--138.
\newblock \href {http://arxiv.org/abs/astro-ph/0105479}
  {\path{arXiv:astro-ph/0105479}}, \href
  {http://dx.doi.org/10.1016/S0370-2693(01)01174-1}
  {\path{doi:10.1016/S0370-2693(01)01174-1}}.

\bibitem{chimento}
L.~P. Chimento, A.~S. Jakubi, D.~Pavon, W.~Zimdahl, {Interacting quintessence
  solution to the coincidence problem}, Phys. Rev. D67 (2003) 083513.
\newblock \href {http://arxiv.org/abs/astro-ph/0303145}
  {\path{arXiv:astro-ph/0303145}}, \href
  {http://dx.doi.org/10.1103/PhysRevD.67.083513}
  {\path{doi:10.1103/PhysRevD.67.083513}}.

\bibitem{delcampo}
S.~del Campo, R.~Herrera, D.~Pavon, {Interacting models may be key to solve the
  cosmic coincidence problem}, JCAP 0901 (2009) 020.
\newblock \href {http://arxiv.org/abs/0812.2210} {\path{arXiv:0812.2210}},
  \href {http://dx.doi.org/10.1088/1475-7516/2009/01/020}
  {\path{doi:10.1088/1475-7516/2009/01/020}}.

\bibitem{cruz}
N.~Cruz, S.~Lepe, F.~Pena, {Dark energy interacting with dark matter and a
  third fluid: Possible EoS for this component}, Phys.Lett. B699 (2011)
  135--140.
\newblock \href {http://dx.doi.org/10.1016/j.physletb.2011.03.049}
  {\path{doi:10.1016/j.physletb.2011.03.049}}.

\bibitem{lepe}
N.~Cruz, S.~Lepe, F.~Pena, {Dark energy interacting with two fluids},
  Phys.Lett. B663 (2008) 338--341.
\newblock \href {http://arxiv.org/abs/0804.3777} {\path{arXiv:0804.3777}},
  \href {http://dx.doi.org/10.1016/j.physletb.2008.04.035}
  {\path{doi:10.1016/j.physletb.2008.04.035}}.

\end{thebibliography}
\end{document}